\documentclass[letterpaper, 10 pt, conference]{ieeeconf}  
\IEEEoverridecommandlockouts                            
\usepackage{graphics} 
\usepackage{epsfig} 
\usepackage{mathptmx} 
\usepackage{times} 
\usepackage{amsmath} 
\usepackage{amssymb}  

\usepackage{array}
\usepackage{stfloats}
\usepackage{url}
\usepackage{verbatim}
\usepackage{cite}
\usepackage{multirow}
\newcolumntype{P}[1]{>{\centering\arraybackslash}p{#1}}
\newcolumntype{M}[1]{>{\centering\arraybackslash}m{#1}}
\usepackage{algorithm}
\usepackage{caption}
\usepackage{subcaption}

\title{\LARGE \bf
Prediction of Neonatal Respiratory Distress in Term Babies\\ at Birth from Digital Stethoscope Recorded Chest Sounds
}

\author{Ethan Grooby$^{1,2}$, \IEEEmembership{Student Member, IEEE},
Chiranjibi Sitaula$^{1}$,
Kenneth Tan$^{3}$,
\\Lindsay Zhou$^{3}$,
Arrabella King$^{3}$,
Ashwin Ramanathan$^{3}$,
Atul Malhotra$^{3}$,
\\Guy A. Dumont$^{2,4}$, \IEEEmembership{Life Fellow, IEEE}, and
Faezeh Marzbanrad$^{1}$, \IEEEmembership{Senior Member, IEEE}
\thanks{E. Grooby  acknowledges the support of the MIME-Monash Partners-CSIRO sponsored PhD research support program and Research Training Program (RTP). A. Malhotra research is supported by the Kathleen Tinsley Trust and a Cerebral Palsy Alliance Research Grant. The study is supported by Monash Institute of Medical Engineering (MIME).}
\thanks{$^{1}$Department of Electrical and Computer Systems Engineering, Monash University, Melbourne, VIC, Australia.}
\thanks{$^{2}$Department of Electrical and Computer Engineering, University British Columbia, Vancouver, BC, Canada}        
\thanks{$^{3}$Monash Newborn, Monash Children’s Hospital and Department of Paediatrics, Monash University, Melbourne, Australia.} 
\thanks{$^{4}$BC Children's Hospital Research Institute, Vancouver, BC, Canada.}     
\thanks{email: ethan.grooby@monash.edu}}      

\begin{document}

\maketitle

\thispagestyle{empty}
\pagestyle{empty}

\begin{abstract}
Neonatal respiratory distress is a common condition that if left untreated, can lead to short- and long-term complications. This paper investigates the usage of digital stethoscope recorded chest sounds taken within 1\,min post-delivery, to enable early detection and prediction of neonatal respiratory distress. Fifty-one term newborns were included in this study, 9 of whom developed respiratory distress. For each newborn, 1\,min anterior and posterior recordings were taken. These recordings were pre-processed to remove noisy segments and obtain high-quality heart and lung sounds. The random undersampling boosting (RUSBoost) classifier was then trained on a variety of features, such as power and vital sign features extracted from the heart and lung sounds. The RUSBoost algorithm produced specificity, sensitivity, and accuracy results of 85.0\%, 66.7\% and 81.8\%, respectively. 
\newline

\indent \textit{Clinical relevance}— This paper investigates the feasibility of digital stethoscope recorded chest sounds for early detection of respiratory distress in term newborn babies, to enable timely treatment and management.
\end{abstract}

\section{Introduction}

Neonatal respiratory distress (RD) is a common condition that affects 5-7\% of term newborns \cite{chowdhury2019full}. This condition is characterised by increased work of breathing which may include: tachypnoea (breathing rate $>60$), nasal flaring, grunting, chest retractions, hypoxemia and/or cyanosis \cite{chowdhury2019full,ramanathan2020assessment}. Early detection of RD is critical to enable clinicians to determine and treat the underlying cause \cite{chowdhury2019full}. Delayed recognition and treatment of RD can lead to both short- and long-term complications such as chronic lung disease, respiratory failure and even death \cite{tochie2016neonatal,chowdhury2019full}. 

Common causes of RD include transient tachypnoea, meconium aspiration syndrome, neonatal pneumonia, respiratory distress syndrome due to surfactant deficiency and pneumothorax \cite{chowdhury2019full}. However, RD can also be the first manifestation of serious, life-threatening conditions such as sepsis, congenital malformations, encephalopathy and metabolic disease \cite{tochie2016neonatal,chowdhury2019full}. Through early detection, clinicians can provide early management such as supportive respiratory care and initiate further investigation for differential diagnosis, in particular, diagnosis of life-threatening conditions \cite{chowdhury2019full}.

For adults, physiological time-series information of respiratory rate, heart rate and oxygen saturation have been used for early detection of RD. With these physiological variables, Markov, convolutional neural network and long short-term memory models have been developed \cite{ravishankar2014early,pardasani2020quantitative,pardasani2020machine}. For newborns, Tochie et al. analysed hospital files to determine important predictors. It was found that acute fetal distress, elective caesarean delivery, APGAR score $<7$ at 1\,min post-delivery, prematurity, male gender and macrosomia were independent predictors of neonatal RD \cite{tochie2016neonatal}. 

Past work has typically relied on continuous monitoring using electrocardiogram and/or pulse oximeter, and detailed hospital files for early detection of RD \cite{ravishankar2014early,pardasani2020quantitative,pardasani2020machine,tochie2016neonatal}. However, in low-resource settings such as rural, developing-world and home environments, this level of monitoring is difficult to achieve. Hence, our work aims to provide accessible RD detection and monitoring through the usage of a digital stethoscope.  

In our previous work, breath sound characteristics at 1\,min post-delivery of all infants who developed RD at some point over the first few hours of life, were compared to those who did not \cite{ramanathan2020assessment}. It was found that there are distinct differences in the frequency sound spectrum. Notably, mean frequency, first-quartile and median frequency were significantly higher in infants with RD than those without. Additionally, the power ratio of 100-200\,Hz was lower and the power ratio of 400-800\,Hz was higher in infants with RD than those without. 

This paper investigates the usage of neonatal chest sound recordings, obtained using a digital stethoscope, to enable early RD detection. Key contributions of this paper are the development of a RD classifier, and usage of both heart and lung sounds individually, as opposed to chest sounds for early RD detection.


\section{Methods}
\label{sec:methods}
\subsection{Data Acquisition}
The study was conducted at Monash Newborn, Monash Children’s Hospital. It was approved by the Monash Health Human Research Ethics Committee (HREA/18/MonH/471). Recordings were obtained 1\,min post-delivery from the right anterior and posterior chest of term newborns using a digital stethoscope. In total, two 1\,min long recordings were obtained for each infant \cite{ramanathan2020assessment}.

These infants were then tracked to check if RD developed in the next few hours of life. For this study, RD was defined as having at least one of the following: breathing rate greater than 60 breaths per minute, nasal flaring, intercostal/subcostal retractions or grunting often requiring supplemental oxygen \cite{ramanathan2020assessment}. In total, 9 infants developed RD and 42 infants did not. 

\subsection{Preprocessing}
A total of 102 60\,s recordings from 51 term newborns were obtained using a digital stethoscope at 44.1\,kHz sampling frequency. These recordings were then low-pass filtered to avoid aliasing and down-sampled to 4kHz.

The chest sounds recordings are the combination of heart, lung and noise sounds. In particular, cry noise and stethoscope movement noise were present in the recordings, overall reducing the quality of the heart and lung sounds. To remove these noise sounds and obtain high-quality heart and lung sounds, we implemented our denoising and sound separation method developed previously in our research \cite{grooby2021new,grooby2022noisy}. 
The method implements non-negative co-factorisation (NMCF) with a reference database of clean heart, lung and noise sounds, to separate the noisy chest sound recording into heart, lung, cry noise, stethoscope movement noise and other noise components \cite{grooby2021new,grooby2022noisy}. 
The noisy chest sound recording was represented in the time-frequency domain using short-time Fourier transfer with a window size of 512 samples, hop size of 256 samples and fast Fourier transform size of 1024 points. For sound separation using NMCF, Kullback-Leibler divergence with sparsity of 0.1 was used in the cost function and the number of basis components were 20, 20, 20, 20 and 10 for heart, lung, cry noise, stethoscope movement noise and unsupervised noise, respectively \cite{grooby2022noisy}. 

After sound separation, there were still segments of the separated heart and lung sound that are of low quality, which do not provide useful information. Using previously developed automated heart and lung signal quality assessment, the separated heart and lung sounds were scored from 1 to 5 based on signal quality, using a sliding window of size 10\,s and hop size of 5\,s \cite{grooby2020neonatal,grooby2021real}. A signal quality score of 1 refers to mostly noise and little or no heart or lung sounds and a signal quality score of 5 refers to clear heart or lung sounds with little or no noise \cite{grooby2020neonatal,grooby2021real}. Signal quality below 3, is considered to be predominantly noise sounds, that are not of diagnostic value \cite{grooby2020neonatal,grooby2021real}. Hence, recording segments with a signal quality of less than 3, were removed. The remaining segments were then concatenated together.

\subsection{Feature Extraction and Selection}
\label{sec:features}
Once clean heart and lung sounds were obtained, a variety of features as presented in Table~\ref{tab1:features} were extracted. Three scenarios were then considered for RD prediction: 
\begin{enumerate}
\item Only anterior chest recordings
\item Only posterior chest recordings
\item Anterior and posterior chest recordings combined
\end{enumerate}

For combining the anterior and posterior chest recordings, the features from the same subject corresponding to anterior and posterior were averaged together. Results for these three scenarios can be seen in Figure~\ref{fig:results}.

For feature ranking, the maximum Relevance Minimum Redundancy (mRMR) algorithm with the mutual information quotient method was used \cite{peng2005feature}. Due to the small dataset, only a subset of features was used to minimise over-fitting. Figure~\ref{fig:feature_rank} shows classification balanced accuracy (average of specificity and sensitivity) results for top 1 to 20 features based on mRMR feature ranking. Based on Figure~\ref{fig:feature_rank} results, the top 19, 1 and 12 features were used for anterior, posterior and combined anterior and posterior scenarios for classification, respectively. The description of these top features is stated in Section~\ref{sec:results}.  

\begin{table}[h]
\caption{The feature set extracted from heart and lung sounds}
    \centering
    \begin{tabular}{|p{1.9cm}||p{5.7cm}|}
    \hline
    \textbf{Title}
    & \textbf{Description}
    \\
    \hline
    Statistical Features
    & Variance, skewness and kurtosis of audio and autocorrelation signal 
    \\
    \hline
    Entropy
    & Sample, Shannon, Renyi and Tsallis entropy 
    \\ 
    \hline
    Power Features
    & Power spectrum represented in dB/octave and intercept and slope of linear regression line calculated. Power spectrum fitted using a 4-term Gaussian mixture model, and the number of peaks, frequency of peaks and 2 highest peaks frequency difference calculated from this. 
    
    Total power, various power ratios from 100-1000Hz, 3dB bandwidth, 1st, 2nd and 3rd quartile, interquartile range, standard deviation, mean frequency, power centroid and max power of power spectrum were calculated. 
    \\
    \hline
    Mel-Frequency Cepstral Coefficients (MFCCs)
    & 13 level decomposition in Mel filter scale and log energy were calculated using a window length of 25\,ms and overlap length of 15\,ms, then minimum, maximum, mean, median, mode, variance and skewness of these frames were calculated.
    \\
    \hline
    Vital Sign Features
    & Heart rate and breathing rate estimated using 10\,s sliding window with 1\,s hop size and then variability calculated \cite{grooby2020neonatal,grooby2021real}. 
    \\
    \hline
    Abnormal Chest Sound Features
    & Using YAMNet, a deep convolutional neural network for sound classification, the probability of groan, grunt, wheeze, gasp, pant, cough and throat clearing sounds present in each from the audio recording was determined \cite{hershey2017cnn,chakraborty2020feature}. 
    \\
    \hline
    Autocorrelation Features
    & Correlation prominence, sinusoid correlation and Hjorth activity to measure the strength of periodicity of the signal \cite{grooby2020neonatal}. 
    \\
    \hline
    \end{tabular}
    \label{tab1:features}
\end{table}

\subsection{Early Respiratory Distress Detection}
Subject-wise cross-validation was performed. During each fold, the training set features were normalised to have zero mean and unit variance. These same scaling and shifting values were then used on the test set features. 

To deal with the imbalanced dataset, the misclassification cost matrix was scaled to balance the dataset. Additionally, a random undersampling boosting (RUSBoost) classifier was used \cite{seiffert2008rusboost}. RUSBoost is an effective classifier for imbalanced datasets by using the combination of undersampling and boosting. Boosting is achieved using the AdaBoost method, which iteratively builds an ensemble of weak learners to create an accurate model \cite{seiffert2008rusboost,freund1997decision}. Each weak learner is trained on the full set of examples from the minority class (RD subjects), and a subset of examples from the majority class (control subjects), through random undersampling \cite{seiffert2008rusboost}. The subset of examples from the majority class differs for each weak learner.   

For our application, decision tree classifiers were used as the weak learners. Using 5-fold cross-validation with Bayesian optimisation, the following parameters of RUSBoost were optimised on the training set: 

\begin{itemize}
\item Number of Weak Learners= 10 to 500
\item Learning Rate= 0.001 to 1
\end{itemize}

\section{Results}
\label{sec:results}

\begin{figure}
     \centering
     \begin{subfigure}[b]{0.48\textwidth}
         \centering
         \includegraphics[scale=0.3,trim={0.5cm 0.5cm 0.5cm 0.5cm}]{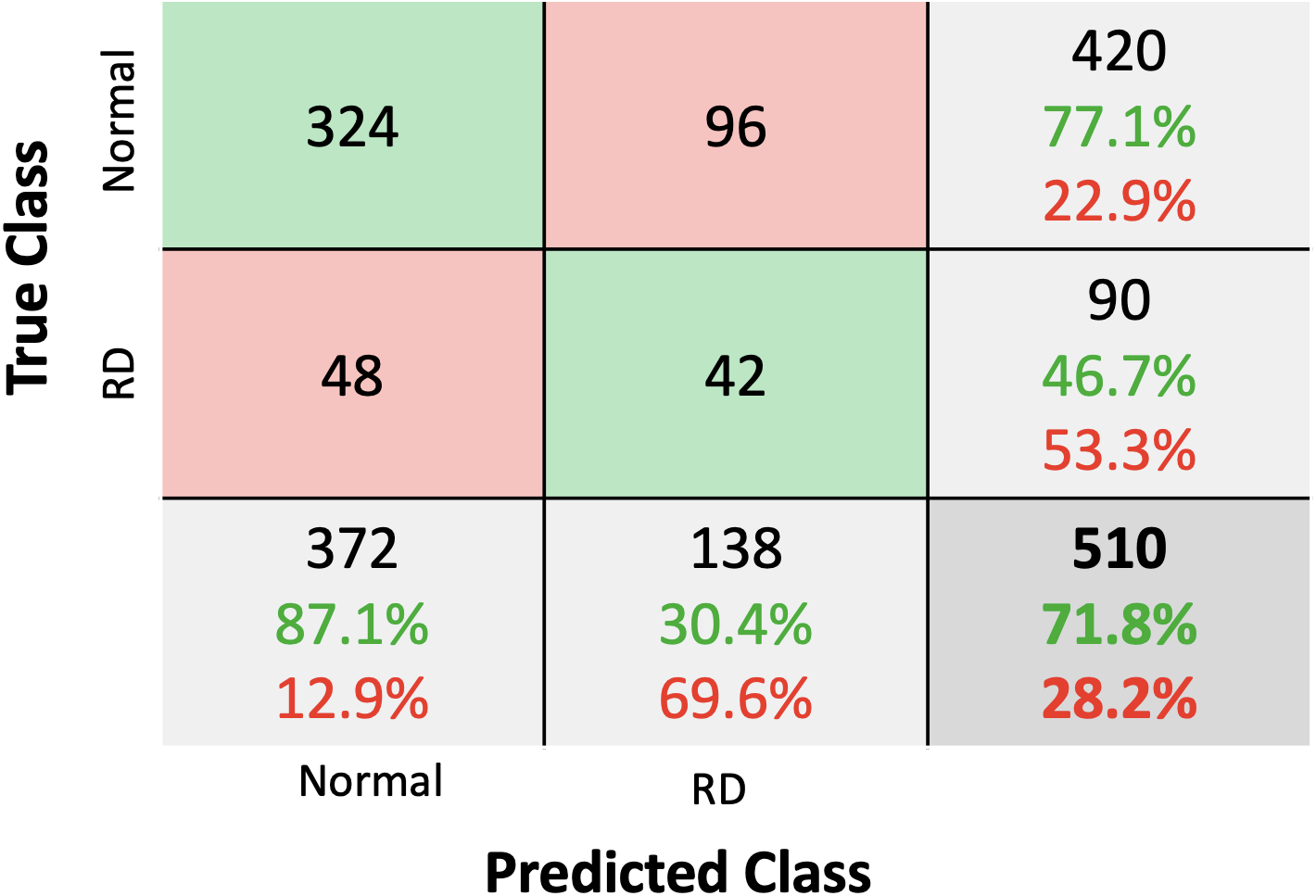}
         \caption{Anterior}
     \end{subfigure}
     \begin{subfigure}[b]{0.48\textwidth}
         \centering
         \includegraphics[scale=0.3,trim={0.5cm 0.5cm 0.5cm 0cm}]{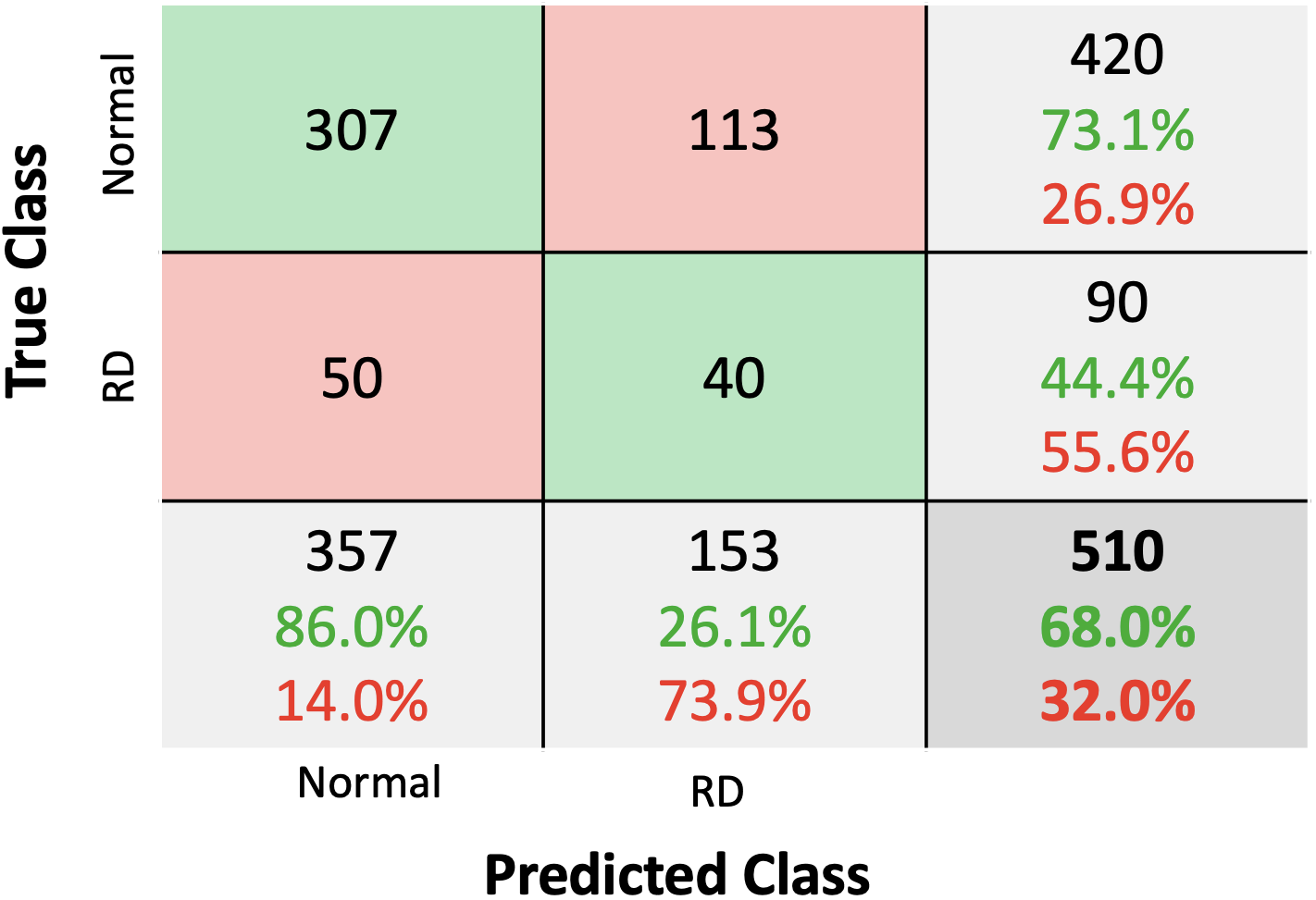}
         \caption{Posterior}
     \end{subfigure}
     \begin{subfigure}[b]{0.48\textwidth}
         \centering
         \includegraphics[scale=0.3,trim={0.5cm 0.5cm 0.5cm 0cm}]{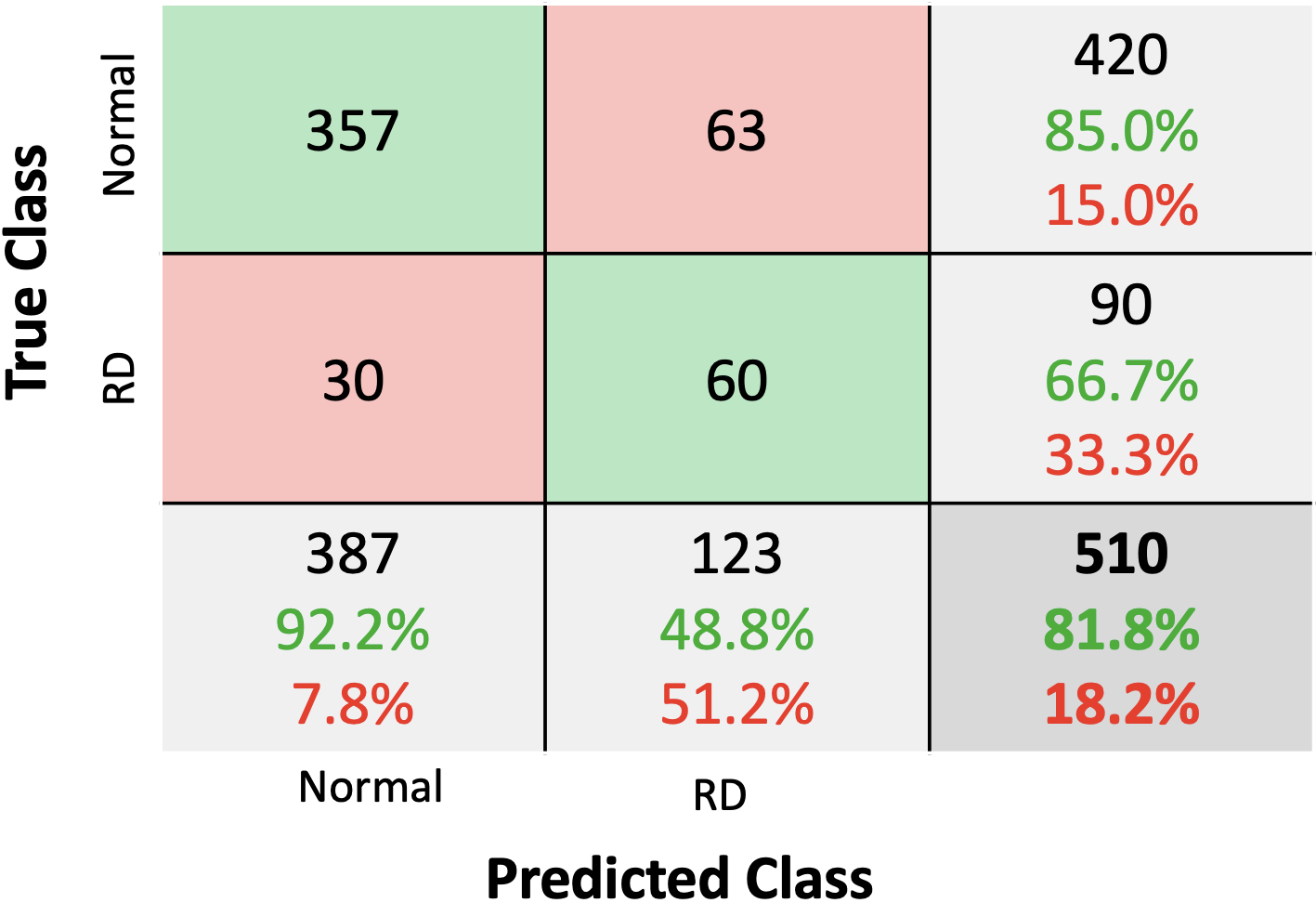}
         \caption{Anterior and Posterior}
     \end{subfigure}
     
        \caption{Confusion matrix of early respiratory distress detection for a) anterior; b) posterior; and c) anterior and posterior chest recordings. The x-axis represents the predicted respiratory distress and the y-axis represents if the infant actually developed respiratory distress. Green and red squares show the number of recordings that have been correctly and incorrectly predicted for respiratory distress, respectively. Grey squares at the end of each row show sensitivity results ($\frac{\text{True Positive}}{\text{True Positive + False Negative}}$). Grey squares at the end of each column show precision results ($\frac{\text{True Positive}}{\text{True Positive + False Positive}}$). Dark grey square summarises the information with total subjects (x 10 iterations) and overall accuracy.\\
        *RD=Respiratory Distress}
        \label{fig:results}
\end{figure}

\begin{figure}
    \centering
    \includegraphics[scale=0.37,trim={0.5cm 0.5cm 1.2cm 0.5cm}]{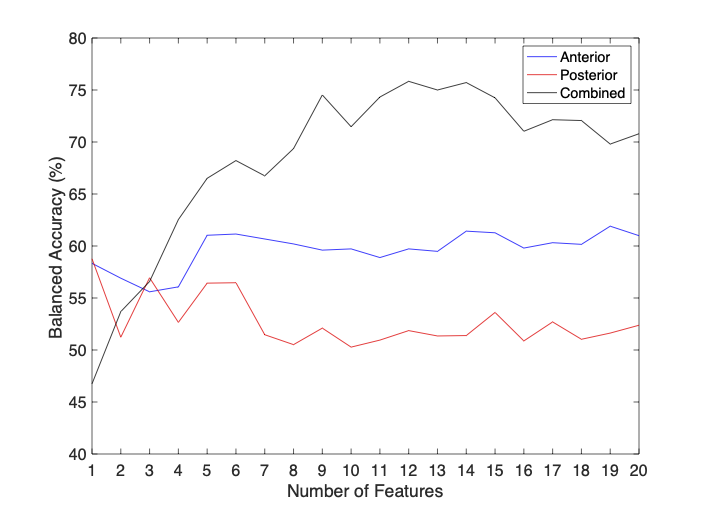}
    \caption{RUSBoost classifier performance for the different number of top features. Balanced accuracy is the average of sensitivity and specificity results. Balance accuracy results are averaged over test subjects in leave-one-subject-out for 10 iterations. Combined refers to combining anterior and posterior features together. Feature selection was performed using the mRMR algorithm as stated in Section~\ref{sec:features}.}
    \label{fig:feature_rank}
\end{figure}

As there was a slight variation in results during each run, the total results for 10 iterations of subject-wise cross-validation are shown in Figures~\ref{fig:results} and \ref{fig:feature_rank}. 

As shown in Figure~\ref{fig:feature_rank}, a combination of anterior and posterior chest recordings produced consistently better balanced accuracy results, once at least 3 features are used. Minimal or no improvement in results is seen past 20 features. 

The confusion matrices of early neonatal RD detection results are presented in Figure~\ref{fig:results}. Using only anterior chest recordings produced specificity, sensitivity and accuracy results of 77.1\%, 46.7\% and 71.8\%. Whereas, only posterior chest recordings produced specificity, sensitivity and accuracy results of 7310\%, 44.4\% and 68.0\%. Combining both anterior and posterior chest recordings together produced an improvement in all results, with specificity, sensitivity and accuracy results of 85.0\%, 66.7\% and 81.8\%. 

Top features based on the mRMR algorithm came from both heart and lung sounds. For lung sounds, MFCC properties related to 867-1071\,Hz frequency range, regression slope line estimation of the power spectrum, log energy and breathing rate were the top features. For heart sounds, Gaussian mixture model fitting parameters and heart rate variability were the top features.

\section{Discussion}
\label{sec:discussion}

Consistent with past works, both lung sound properties and vital signs (heart and breathing rate) are useful in early RD detection \cite{ravishankar2014early,pardasani2020quantitative,pardasani2020machine}. As the autonomic nervous system regulates both heart and breathing rate, variability in these vital signs can suggest this is the underlying cause of sickness. Additionally, it was found that there are distinct power spectrum properties, not only in lung sounds but also heart sounds in newborns that develop RD. Further research would be required to understand the reason, but there are several possibilities. Firstly, the underlying cause of RD may be cardiac-related, hence the distinct frequency properties \cite{chowdhury2019full}. Secondly, increased work of breathing associated with RD is also correlated with the increased cardiovascular workload. 

As seen in Figure~\ref{fig:results}, anterior recordings outperformed posterior recordings. This can be explained by the fact anterior recordings typically contain both strong heart and lung sounds, which were required to extract useful features. Whereas, posterior recordings have strong lung sounds and weak heart sounds. This is exemplified by the fact 59.3\% of heart sound segments were removed due to poor quality in the posterior recording set, as opposed to 40.0\% in the anterior recording set. However, the benefit of posterior recordings is additional lung sound information and higher quality lung sounds, with only 27.0\% of lung sound segments being removed due to poor quality in comparison to  32.5\% in the anterior recording set. Overall, the combination of both anterior and posterior recordings enables more accurate and detailed heart and lung sound features, resulting in improved classifier performance.  

As stated in the previous paragraph, a large percentage of segments were removed due to poor quality. This suggests that future work is required in the denoising and sound separation method, to enable not only high-quality heart and lung sounds, but a larger amount of information to work with. Additionally, longer recordings may improve results as there is a higher likelihood of getting clean segments. 

\section{Conclusion}
In this paper, we used both heart and lung sounds extracted from chest sound recordings for RD detection. We found that the combination of both anterior and posterior chest recordings produces promising results with an overall accuracy of 81.8\% for RD detection, outperforming individual anterior and posterior recordings. This underscores the utility of stethoscope-recorded chest sounds for early RD detection in newborns. However, this is based on a small dataset and a larger study would be required to make more definitive conclusions.
\label{sec:conclusion}

\bibliographystyle{IEEEtran}
\bibliography{IEEEabrv,references}
\end{document}